\documentclass{llncs}
\usepackage{makeidx}  

\usepackage{cite}
\usepackage{latexsym}
\usepackage{amsmath}
\usepackage{amsfonts}
\usepackage{epsfig}
\usepackage{epsf}
\usepackage{latexsym}
\usepackage{url}
\usepackage{wrapfig}
\usepackage{subfigure}
\usepackage{indentfirst}
\usepackage{graphics}
\usepackage{graphicx}
\usepackage{verbatim}
\usepackage{color}

  \let\[=\(
  \let\]=\)
\newenvironment{boxalg}[1]{{\vspace{1em}}{\noindent}\fbox{\begin{minipage}{0.985\linewidth}
                        \vspace{1em}
                        \makebox[0.025\linewidth]{}
                        \begin{minipage}{0.95\linewidth}
                        #1
                        \end{minipage}
                        \end{minipage}}{\vspace{1em}}}

\newcommand{\sketch}[1]{{\noindent {\it Sketch of Proof.} {#1} \qed \vspace{1ex}}} 

\newcommand{\eps}{\varepsilon}
\newcommand{\ol}[1]{\overline{#1}}
\newcommand{\OPT}{\mbox{\it OPT}}
\newcommand{\ALG}{\mbox{\it MM}}
\newcommand{\MM}{\mbox{\it MM}}
\newcommand{\HEU}{\mbox{\it MM}}
\newcommand{\MC}{\mbox{\it MC1x1}}
\newcommand{\OPTij}{\mbox{\it OPT}_{ij}}
\newcommand{\OPTip}{\mbox{\it OPT}_{i\bullet}}

\newcommand{\OPTpj}{\mbox{\it OPT}_{\bullet j}}

\newcommand{\notOPTip}{\ol{\mbox{\it OPT}}_{i\bullet}}

\newcommand{\notOPTpj}{\ol{\mbox{\it OPT}}_{\bullet j}}
\newcommand{\HEUij}{\mbox{\it MM}_{ij}}
\newcommand{\HEUip}{\mbox{\it MM}_{i\bullet}}
\newcommand{\HEUpj}{\mbox{\it MM}_{\bullet j}}
\newcommand{\HEUop}{\mbox{\it MM}_{0\bullet}}

\newcommand{\notHEUip}{\ol{\mbox{\it MM}}_{i\bullet}}
\newcommand{\notHEUpj}{\ol{\mbox{\it MM}}_{\bullet j}}
\newcommand{\notHEUop}{\ol{\mbox{\it MM}}_{0\bullet}}

\begin{document}
\mainmatter

\title{Communication-Aware Processor Allocation for Supercomputers}
%\thanks{
%Extended Abstract. A full version is available as \cite{bdd+-capas-04}.}}

\author{Michael A. Bender\inst{1} \and
David P. Bunde\inst{2} \and
Erik D. Demaine\inst{3} \and
S\'andor P. Fekete\inst{4} \and
Vitus J. Leung\inst{5} \and
Henk Meijer\inst{6}
\and
Cynthia A. Phillips\inst{5}
}
\institute{
Department of Computer Science,\\
    SUNY Stony Brook, Stony Brook, NY 11794-4400, USA.\\
    \email{bender@cs.sunysb.edu}.\\
 \and
Department of Computer Science,\\
University of Illinois, Urbana, IL 61801, USA.\\
  \email{bunde@uiuc.edu}.\\
  \and
   MIT Computer Science and Artificial Intelligence Laboratory,\\
   Cambridge, MA 02139, USA.\\
   \email{edemaine@mit.edu}.
\and
    Dept.\ of Mathematical Optimization,\\
            Braunschweig University of Technology,\\
            38106 Braunschweig, Germany.\\
            \email{s.fekete@tu-bs.de}.\\
  \and
Discrete Algorithms \& Math Department,\\
Sandia National Laboratories, 
Albuquerque, NM 87185-1110, USA.\\
\{\texttt{vjleung}, \texttt{caphill}\}\texttt{@sandia.gov}.
 \and
Dept.\ of Computing and Information Science,\\
Queen's University,\\
Kingston, Ontario, K7L 3N6, Canada.\\
            \email{henk@cs.queensu.ca}.
}

\bibliographystyle{abbrv}
\maketitle
\thispagestyle{plain}

\begin{abstract}
We give processor-allocation algorithms for grid architectures,
where the objective is to select processors from a set of available
processors to minimize the average number of communication hops.

The associated clustering problem is as follows: Given $n$ points in
$\Re^d$, find a size-$k$ subset with minimum average pairwise $L_1$
distance. We present a natural approximation algorithm and show that it is a
$\frac74$-approximation for 2D grids. In $d$ dimensions,
the approximation guarantee is $2-\frac{1}{2d}$,
which is tight. We also give a polynomial-time
approximation scheme (PTAS) for constant dimension $d$
and report on experimental results.
\end{abstract}

\section{Introduction}
\label{sec:intro}

We give processor-allocation algorithms for grid architectures.  Our
objective is to select processors to run a job from a set of
available processors so that the average number of communication
hops between processors assigned to the job is minimized.
Our problem is restated as follows:
given a set $P$ of $n$ points in $\Re^d$, find a
subset $S$ of $k$ points with minimum average pairwise $L_1$ distance.

\paragraph{Motivation: Processor Allocation in Supercomputers.}
Our algorithmic work is motivated by a problem in the operation of
supercomputers. 
Specifically, we targeted our algorithms and simulations at
Cplant~\cite{brightwell00massively,cplant}, a commodity-based
supercomputer developed at Sandia National Laboratories, and Red Storm, a
custom supercomputer being developed at Cray, though other
supercomputers at Sandia have similar features.
In these systems, a scheduler selects the next job to run based on priority.
The allocator then independently places the job on a set of processors
which exclusively run that job to completion.  Security constraints
forbid migration, preemption, or multitasking.
These constraints make the allocation decision more important
since it cannot be changed once made.

To obtain maximum
throughput in a network-limited computing system, the processors
allocated to a single job should be physically near each other. This
placement reduces communication costs and avoids bandwidth contention
caused by overlapping jobs. 
Experiments have shown that allocating nearby processors to each job
can improve throughput on a range of
architectures~\cite{baylor96,leung02a,mache96,mache97b,moore96}.  
Several papers suggest that minimizing the \emph{average number of
communication hops} is an appropriate metric for job
placement~\cite{mache96,mache97b,krumke97}.  Experiments with a
communication test suite demonstrate that this metric correlates with
a job's completion time~\cite{leung02a}.

Early processor-allocation algorithms allocate only convex sets of
processors to each job~\cite{li91,chuang91,zhu92,bhattacharya94}.
For such allocations,
each job's communication can be routed entirely within processors assigned to that job, so
jobs contend only with themselves.  But requiring convex allocations
reduces the achievable system utilization to levels unacceptable for
a government-audited system~\cite{krueger94,subramani02}.

%\vspace*{-3mm}
\begin{figure}
\begin{center}
\includegraphics[width=0.7\textwidth]{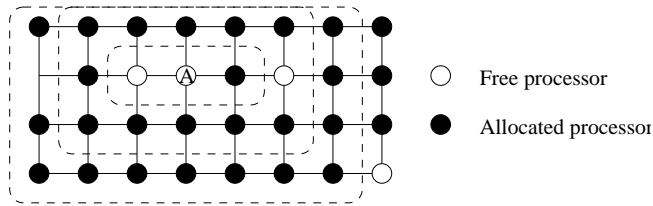}
\caption{Illustration of MC: Shells
around processor $A$ for a $3 \times 1$ request.}
\label{mc-figure}
\end{center}
\end{figure}

%\vspace*{-4mm}
Recent work~\cite{lo97,mache97a,chang98,leung02a,subramani02} allows
discontiguous allocation of processors but tries to cluster them and
minimize contention with previously allocated jobs.  Mache, Lo, and
Windisch~\cite{mache97a} propose the MC algorithm for grid
architectures:
For each free processor, algorithm MC evaluates the quality of an
allocation centered on that processor. It counts the number of free
processors within a submesh of the requested size centered on the
given processor and within ``shells'' of processors around this
submesh; see Figure~\ref{mc-figure} reproduced
from~\cite{mache97a}.
The cost of an allocation is the sum of the shell numbers of the
allocated processors.
MC chooses the allocation with lowest cost.
Since users at Sandia do not request processors in a particular shape,
in this paper, we consider MC1x1, a variant in which
shell 0 is $1 \times 1$ and subsequent
shells grow in the same way as in MC.

Originally, processor allocation on the Cplant system was
\emph{not} based on the locations of the free processors.  The
allocator simply verified that enough processors were
free before dispatching a job.  The current allocator uses space-filling
curves and 1D bin-packing techniques based upon work
of Leung et al.~\cite{leung02a}.
We also have Cplant implementations of a 3D version of MC1x1 and the
greedy heuristic (called MM) analyzed in this paper.

\paragraph{Related Algorithmic Work.}
Krumke et al.~\cite{krumke97} consider a generalization of our problem on
arbitrary topologies for several measures of
locality, motivated by allocation on the CM5.
They prove it is NP-hard to approximate average pairwise distance
in general, but give a 2-approximation for distances obeying the triangle
inequality.

\begin{figure}[hbt]
\centering{
\includegraphics[width=.6\textwidth]{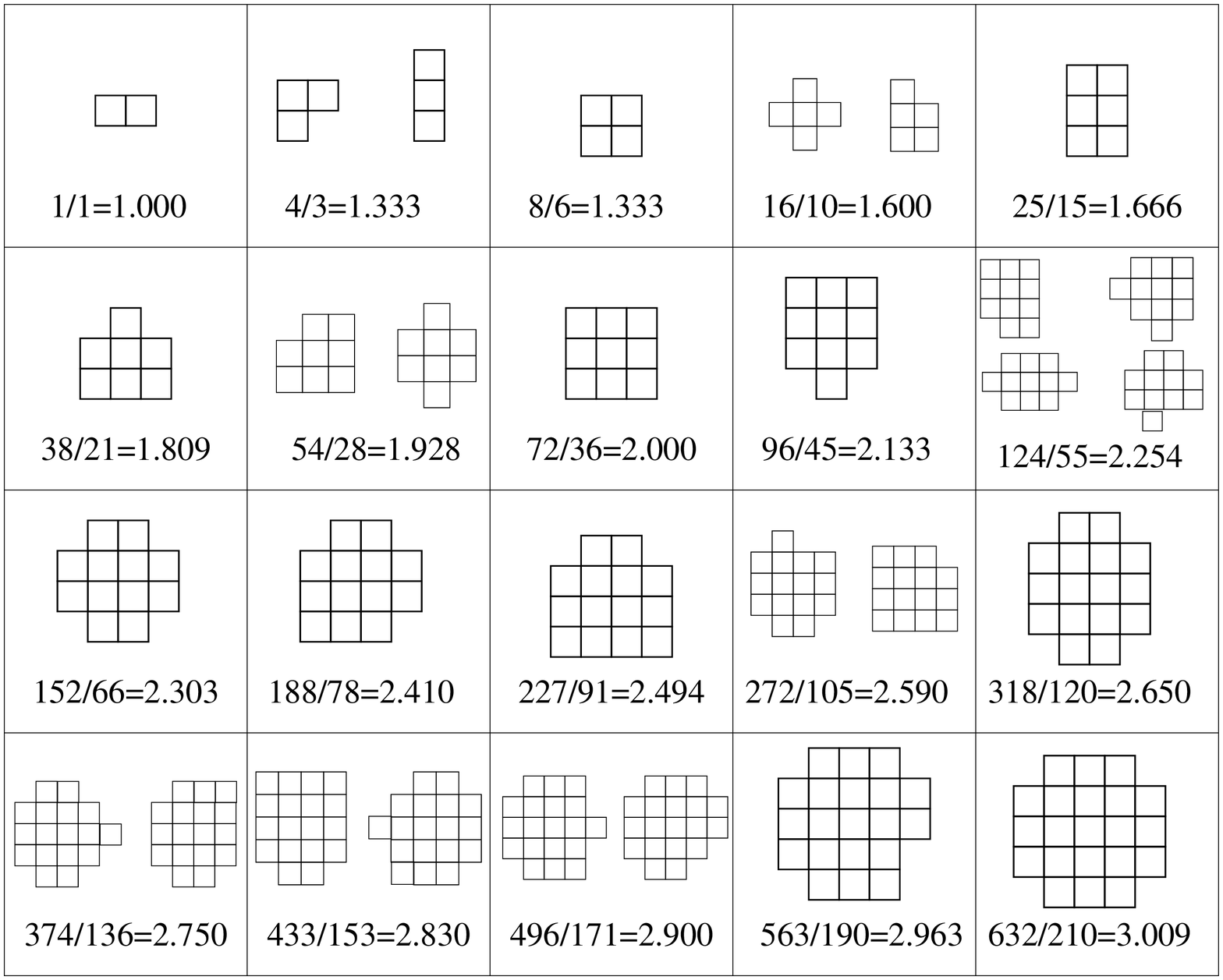}}
\caption{\small Optimal unconstrained clusters for small
values of $k$; numbers shown are the average $L_1$ distances, with
truncated decimal values.} \label{fi:disc-town}
\end{figure}

\begin{figure}[hbt]
\centering{
\includegraphics[width=.6\textwidth]{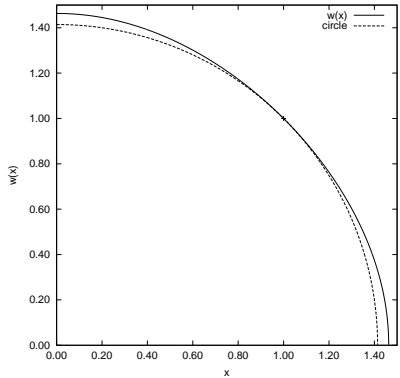}}
\caption{\small 
Plot from Bender et al.~\cite{BenderBeDeFe02} of
a quarter of the optimal limiting boundary curve for the unconstrained
problem; the dotted line is
a circle.} \label{fi:cont-town}
\end{figure}

A natural special case of the allocation problem is the
{\em unconstrained\/} problem, in the absence of occupied processors:
For any
number $k$, find $k$ grid points minimizing average pairwise
$L_1$ distance.  For moderate values of $k$, these sets can
be found by exhaustive search; see Figure~\ref{fi:disc-town}.  The
resulting shapes appear to approximate some ``ideal'' rounded shape,
with better and better approximation for growing $k$.  Karp et
al.~\cite{KarpMcWo75} and Bender et al.~\cite{BenderBeDeFe02} study
the exact nature of this shape, shown in Figure~\ref{fi:cont-town}.  
Surprisingly, there is no known closed-form solution for the resulting
convex curve, but Bender et al.~\cite{BenderBeDeFe02} have 
expressed it as a differential equation.
The complexity of this special case remains open, but its
mathematical difficulty suggests the hardness of obtaining
good solutions for the general constrained  problem.

%\vspace*{-6mm}

%\vspace*{-6mm}
In reconfigurable computing on field-programmable gate arrays
(FPGAs), varying processor sizes give rise to a generalization of
our problem:  place a set of rectangular modules on a grid to
minimize the overall weighted sum of $L_1$ distances between
modules. Ahmadinia et al.~\cite{abftv-orcdprd-04} give an optimal
$\Theta(n\log n)$ algorithm for finding an optimal feasible location
for a module given a set of $n$ existing modules. At this point, no
results are known for the general off-line problem (place $n$
modules simultaneously) or for on-line versions.

Another related problem is \emph{min-sum $k$-clustering}:
separate a graph into $k$ clusters to minimize the sum of distances
between nodes in the same cluster.  For general graphs, Sahni and
Gonzalez~\cite{sahni76} show it is NP-hard to approximate this
problem to within any constant factor for $k\geq 3$. In a metric
space,
Guttmann-Beck and
Hassin~\cite{guttmannbeck98} give a $2$-approximation,
Indyk~\cite{indyk99a} gives a PTAS for $k=2$, and Bartel et
al.~\cite{bartal01} give an
$O((1/\epsilon)\log^{1+\epsilon} n)$-approximation for general $k$.

Fekete and Meijer~\cite{fm-mdgmwc-04} consider the problem of {\em
maximizing} the average $L_1$ distance. They give a PTAS for this
{\em dispersion\/} problem in $\Re^d$ for constant $d$, and show
that an optimal set of any fixed size can be found in $O(n)$ time.

\paragraph{Our Results.}
\label{sec:results}

We consider algorithms for minimizing the average
$L_1$ distance between allocated processors in a mesh supercomputer.
In particular, we give the following results:

%\vspace*{-1mm}
\begin{itemize}
\item We prove that a greedy algorithm we call MM is a
$\frac74$-approximation algorithm for $2D$ grids.
This reduces the previous best factor of 2~\cite{krumke97}.
We show that this analysis is tight.

\item We present a simple generalization of MM to $d$-dimensional grids
and prove that it gives a $2-\frac{1}{2d}$ approximation,
which is tight.

\item We give a polynomial-time approximation scheme (PTAS) for points
in $\Re^d$ for constant $d$.

\item Using simulations, we compare the allocation performance of MM
to that of other algorithms.
As a byproduct, we get insight on how to place a stream of jobs in
an online setting.

\item
We give an algorithm to exactly solve the 2-dimensional case for $k=3$
in time $O(n\log n)$.

\item
We prove that the $d$-dimensional version of MC1x1 has approximation
factor at most $d$ times that of MM. 

\end{itemize}

Our work also led to a linear-time dynamic programming algorithm for
the 1-dimensional problem of points on a line or ring; see
Leung et al.~\cite{benderBDFL03} for details.

%\section{\sloppy Manhattan Median Algorithm for Two-Dimensional Point Sets}
\section{\sloppy Algorithms for Two-Dimensional Point Sets}

\subsection{Manhattan Median Algorithm}
\label{sec:two-approximation}

Given a set $S$ of $k$ points in the plane, a point that minimizes the
total $L_1$ distance to these points is called an ($L_1$)
median.  Given the nature of $L_1$ distances, this is a point
whose $x$-coordinate (resp. $y$-coordinate) is the median of the $x$
(resp. $y$) values of the given point set.  We can always pick a
median whose coordinates are from the coordinates in $S$.  There is a
unique median if $k$ is odd; if $k$ is even, possible median
coordinates may form intervals.

The natural greedy algorithm for our clustering problem is as follows:

\begin{boxalg}{
%\vspace*{-3mm}
Consider the set $I$ containing the $O(n^2)$ intersection points of the
horizontal and vertical lines through the points of input P.
For each point $p \in I$ do:
\begin{enumerate}

\item Take the $k$ points closest to $p$ (using the $L_1$ metric), breaking ties arbitrarily.

\item Compute the total pairwise distance between all
$k$ points.

\end{enumerate}
Return the set of $k$ points with smallest total pairwise distance.
}

\end{boxalg}

We call this strategy MM, for {\bf M}anhattan {\bf M}edian. We prove
that MM is a $\frac{7}{4}$-approximation on 2D meshes. (Note that
Krumke et al.~\cite{krumke97} call a minor
variation of this algorithm Gen-Alg and show
it is a 2-approximation in arbitrary metric spaces.)

%\subsection{Analysis of the Algorithm}

For $S \subseteq P$, let $|S|$ denote the sum of $L_1$
distances between points in $S$.
For a point $p$ in the plane, we use $p_x$ and $p_y$ to denote
its $x$- and $y$-coordinates respectively.

\begin{lemma}
\label{le:lower}
\ALG\ is not better than a $7/4$ approximation.
\end{lemma}

\begin{figure}[htb]
\begin{center}
  \input{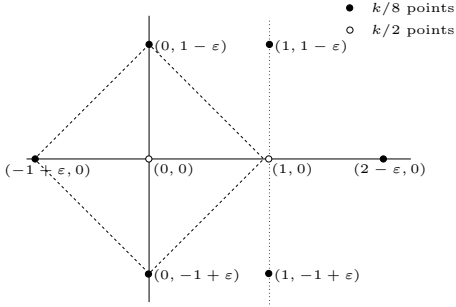}
  \caption{A class of examples where MM yields a ratio of $7/4$.}
  \label{fi:bad}
\end{center}
\end{figure}

\begin{proof}
For a class of examples establishing the lower bound,
consider the situation shown in Figure~\ref{fi:bad}.
For any $\epsilon>0$, it has clusters of $k/2$ points at $(0, 0)$ and
$(1, 0)$.
In addition, it has clusters of $k/8$ points at $(0, \pm(1-\epsilon))$,
$(1, \pm(1-\epsilon))$, $(2-\epsilon, 0)$, and $(-1+\epsilon, 0)$.
The best choices of median are $(0, 0)$ and $(1, 0)$, which yield
a total distance of $7k^2 (1-\Theta(\epsilon))/16$.
The optimal solution is the points at $(0, 0)$ and $(1, 0)$, which
yield a total distance of $k^2/4$.
\qed
\end{proof}

%\vspace*{-8mm}

%\vspace*{-8mm}
Now we show that $7/4$ is indeed the worst-case bound. We focus on
possible worst-case arrangements and use local optimality to
restrict the possible arrangements until the claim follows.

Let $\OPT$ be a subset of $P$ of size $k$ for which $|\OPT|$ is minimum.
Without loss of generality assume that the origin is a median point
of $\OPT$.
This means that at most $k/2$ points of $\OPT$ have positive
$x$-coordinates
(similarly negative $x$-coordinates, positive $y$-coordinates, and negative
$y$-coordinates).
Let $\ALG$ be the  set of $k$ points closest to the origin.
Since this is one candidate solution for the algorithm, its
sum of pairwise distances is at least as high as that of the solution
returned by the algorithm.

Without loss of generality,
assume that the largest $L_1$ distance of a point in $\ALG$ to the origin
is 1, so $\ALG$ lies in the $L_1$ unit circle $C$.
(Note that $C$ is diamond-shaped.)
We say that points are either inside $C$, on $C$, or outside $C$.
All points
of $P$ inside $C$ are in $\ALG$ and at least some points
on $C$ are in $\ALG$.
If there are more than $k$ points on and inside $C$, we select
all points inside $C$ plus those points on $C$
maximizing $|\ALG|$.

Clearly $1 \leq |\MM|/|\OPT|$. 
Let $\rho_k$ be the supremum of $|\ALG|/|\OPT|$ over all inputs $P$.
By assuming that ties are broken badly, we can assume that there is
an input
% configuration $S \subseteq P$ 
for which $|\ALG|/|\OPT| = \rho_k$:

%\vspace*{-2mm}
\begin{lemma}
\label{le:attain}
For any $n$ and $k$, there are point sets $P$
with $|P|=n$ for which $|\ALG|/|\OPT|$ attains
the value $\rho_k$.
\end{lemma}

%\vspace*{-5mm}
\begin{proof}
The set of arrangements of $n$ points in the unit circle $C$ is a
compact set in $2d$-dimensional space. By our assumption on breaking
ties, $|\ALG|/|\OPT|$ is upper semicontinuous, so it attains a
maximum.\qed
\end{proof}

We show that $|\ALG|$ is at most $7/4$ times larger than  $|\OPT|$.

\begin{theorem}
\label{th:7/4}
\MM\ is a $7/4$-approximation algorithm for minimizing the sum of
pairwise $L_1$ distances in a 2D mesh.
\end{theorem}

\begin{proof}
For ease of presentation, we assume without loss of generality
that $P = \ALG \cup \OPT$.
Let $B = \OPT \cap \ALG$,  $O = \OPT-B$ and $A = \ALG-B$.
%See Figure~\ref{fi:claim4}~(a).

{\bf Claim 0:} {\em No point $p \in O$ lies outside $C$.}

If a point $p \in O$ lies outside $C$ we can move it a little closer to
the origin without entering $C$. Since it remains
outside $C$, the point does not become part of $\ALG$,
so $|\OPT|$ is reduced, $|\ALG|$ remains the same and the ratio
$|\ALG|/|\OPT|$ increases, which is impossible.

{\bf Claim 1:} {\em All points inside $C$ are in $\ALG$.}

It follows from the definition of $\ALG$ that all points
inside $C$ are in $\ALG$. Notice that this implies that no point
$p \in O$ can lie inside $C$.

{\bf Claim 2:} {\em Without loss of generality, we may assume that the origin is also a median of $\ALG$.}

Suppose that the origin is not a median of $\ALG$.
We consider the case when more than $k/2$ points of $\ALG$
have positive $y$-coordinate; the other cases are handled
analogously.
%Without loss of generality, assume that
%there are more than $k/2$ points from $\ALG$
%with positive $y$-coordinate.
%So any median of $\ALG$ has a positive $y$-coordinate.
We set the $y$-coordinate of the point in $\ALG$ with smallest
positive $y$-coordinate to zero.
By assumption, this causes the point to move away from at least as
many points of $\ALG$ as it moves toward.
Thus, $|\ALG|$ does not decrease.
The origin is a median of $\OPT$ so $|\OPT|$ does not increase.
Therefore, the ratio $|\ALG|/|\OPT|$ cannot decrease.
Since the ratio cannot increase by assumption, it
must remain the same.
Thus, we have constructed a point set achieving $|\ALG|/|\OPT|=\rho_k$
with one fewer point having positive $y$-coordinate.
Repeating this process will make some point on the line $y=0$ a median.

{\bf Claim 3:} {\em No point $p \in A$ lies inside $C$.}

Suppose there is a $p \in A$ that lies inside $C$.
Moving $p$ away from the origin increases $\ALG$ because
$p$ is moved further away from the median of $\ALG$. Since $p \notin \OPT$,
$\OPT$ does not increase, although it may decrease.
So $|\ALG|/|\OPT|$ increases, which is impossible.
This implies that all points inside $C$ are in $B$
and that points from $A$ and $O$ lie on the boundary of $C$.

{\bf Claim 4:} { \em Without loss of generality, we may assume
that all points $p \in A$ on $C$  lie  in a corner of $C$.}

Suppose $p \in A$ lies on an edge of $C$ but not in a corner. Let
$D$ be the sum of the $L_1$ distances from $p$ to all points in $\ALG-{p}$.
Consider the set $Q$ of all points $q$ for which the sum
of the $L_1$ distances from $q$ to all points in $\ALG-{p}$
is at most $D$.
The sum of distances is the sum of convex functions so it is also a
convex function and the set $Q$ is a convex polygon through $p$.
Therefore, we can move $p$ along the edge of $C$ on which it lies so that
it either moves outside of $Q$ or remains on the boundary of $Q$.
In former case, $|\ALG|$ increases.
In the latter, $|\ALG|$ remains the same.
%equal or possibly increases if $p$ leaves $\ALG$.
In either case, $|\OPT|$ stays the same or decreases.
If $|\ALG|$ increases and/or $|\OPT|$ decreases, $|\ALG|/|\OPT|$ increases
which is impossible.
If both stay the same, we can move $p$ until
it reaches a corner of $C$.
For an illustration of what the configuration may look like
see Figure \ref{fi:claim4}(a).

\begin{figure}[htb]
\begin{center}
  \input{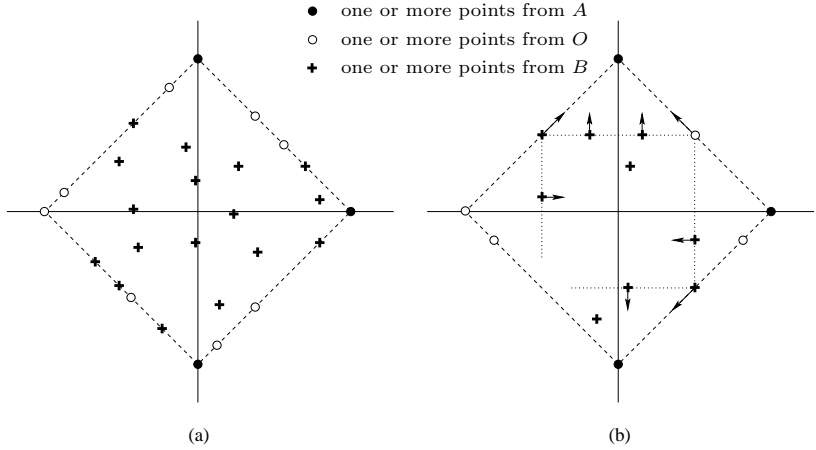}%
  \caption{Points of $A$, $O$ and $B$ (a) after
claim 4 and (b) during motion used in claim 5.}
  \label{fi:claim4}
\end{center}
\end{figure}

{\bf Claim 5:}
{\em Without loss of generality we may assume
that all points in $O \cup B$ lie in a corner of $C$
or on the origin.}

We prove the claim by contradiction. Suppose there is a set of points
$S$ for which the claim is false.
Let $p\in O \cup B$ be a point that does not lie in a corner of $C$
or on the origin.
Let $S(p)$ be
the points that lie
on the axis-parallel rectangle through $p$ with corners on $C$.
% We move these points
% simultaneously, in such a way that they stay on an axis-parallel
% Without loss of generality assume that $0 < p_y < 1$.
% We now define up to four sets of points.
% Let $Y^+(p)$ be the points in $S$ with $y$-coordinate equal to $p_y$.
% If there is a point $q \in Y^+(p)$ with $q_x + q_y = 1$ then let
% $X^+(p)$ be the points in $S$ with $x$-coordinate equal to $q_x$,
% otherwise $X^+(p)$ is empty.
% If there is a point $q \in Y^+(p)$ with $q_y - q_x = 1$ then let
% $X^-(p)$ be the points in $S$ with $x$-coordinate equal to $q_x$.
% If there is a point $q \in X^+(p) \cup X^-(p)$ with
% $q_x + q_y = -1$ or
% $q_x - q_y = 1$ then let
% $Y^-(p)$ be the points in $S$ with $y$-coordinate equal to $q_y$.
% So the four sets of points lie on four axis-parallel line segments in the disk
% that meet on $C$.
% Let $XY(p) = X^-(p) \cup X^+(p) \cup Y^-(p) \cup Y^+(p)$.
The set $S(p)$ is illustrated in  Figure \ref{fi:claim4}(b).
We move the points in $S(p)$
simultaneously in such a way that they stay on
an axis-parallel rectangle with corners on $C$.
For example we move all points in $S(p)$ with maximal $y$-coordinates but
not on $C$ upwards by $\epsilon$.
We move all points in $S(p)$ with maximal $y$-coordinates
and on $C$ upwards while remaining on $C$.
Similarly the other points of $S(p)$ move either left, right or down.
We choose $\epsilon$ small enough such that no point from $S \setminus S(p)$
enters the rectangle on which $S(p)$ lies.
This move changes $|\ALG|$ by some amount $\delta_a$
and $|\OPT|$ by some amount $\delta_o$. However if we move all
points in the opposite direction (i.e. points with maximal $y$-coordinates
downwards, etc.) $|\ALG|$ and $|\OPT|$ change by $-\delta_a$ and $-\delta_o$
respectively.
So if $\delta_a/\delta_o \neq \rho_k$, one of these two moves
increases $|\ALG|/|\OPT|$, which is impossible.
If $\delta_a/\delta_o = \rho_k$ we keep moving the points in the same
direction until there is a combinatorial change, i.e.
a point from $S \setminus S(p)$ enters the rectangle on which $S(p)$ lies, 
a point in $S(p)$ reaches $C$, 
or the rectangle collapses into a line.
Each combinatorial change decreases the number of rectangles on which
the points lie, increases the number of points on $C$, or moves points
to one of the coordinate axes.
Since none of these changes is ever undone, we can then repeat this argument
until all points of $S$ lie on a corner of $C$ or on the origin.

%\vspace*{2mm}
We can now complete the proof of Theorem~\ref{th:7/4}.
Let $b$ denote the number of points at the origin.
These points are all in $B=\OPT\cap\ALG$ since they were originally inside $C$.
Let $a_0,a_1,a_2,a_3$
and $o_0,o_1,o_2,o_3$ be the points of $\ALG$ and $\OPT$
at the north, east, south and west corners of $C$ respectively.
The value of $|\ALG|$ is
\[       2\sum_{0 \leq i < j \leq 3} a_i a_j + \sum_{0 \leq i \leq 3} b a_i
   ~~=~~ 2\sum_{0 \leq i < j \leq 3} a_i a_j + b(k-b) \]
which is maximal when each value $a_i$ is equal to $\lceil (k-b)/4 \rceil$
or $\lfloor (k-b)/4 \rfloor$.
The value of $|\OPT|$ is
\[ 2\sum_{0 \leq i < j \leq 3} o_i o_j +  b(k-b) \]
which is minimal when $o_0 = k-b$ and $o_1 = o_2 = o_3 = 0$.
The origin must be a median of $\OPT$ since none of our
transformations move a point between quadrants.
Thus, if $b < k/2$, the minimum
value for $|\OPT|$ occurs when $o_0 = k/2$ and $o_1 = k/2 - b$.
So if  $b < k/2$ we have
\[ \frac{|\ALG|}{|\OPT|} \leq \frac{\frac{12(k-b)^2}{16} + b(k-b)}
            {k(\frac{k}{2}-b) + b(k-b)}   \]
from which it follows that
\[ \frac{|\ALG|}{|\OPT|} \leq \frac{3k^2 - 2kb - b^2}
                                 {2k^2 - 4b^2}  . \]
This is a convex function of $b$ in the interval $0 \leq b < k/2$
whose values are smaller than 7/4. 

If $b\geq k/2$ we have
\[ \frac{|\ALG|}{|\OPT|} \leq \frac{\frac{12(k-b)^2}{16} + b(k-b)}
        {b(k-b)} ~=~ \frac{3k+b}{4b}     \] 
which is maximal when $b = k/2$ in which case $|\ALG|/|\OPT| = 7/4$.
Notice that $n$ has to be at least $11k/8$ for this value
to be obtained since we need $a_i = k/8$ for all $i$
and $o_0 = k/2$ where $\ALG$ and $\OPT$ can share
the points in the north corner of $C$. 
For smaller values of  $n$
we can add extra points to the corners
of $C$ until $n = 11k/8$, so $\ALG$ increases and $\OPT$ decreases.
Since $|\ALG/|\OPT| = 7/4$ when $n = 11k/8$ we have
$|\ALG/|\OPT|  \leq 7/4$  for all values of $k$.
Therefore the theorem holds.
\qed
\end{proof}

\subsection{Analysis of MC1x1}

MC was originally presented as a heuristic algorithm, but we prove
that MC1x1 has approximation ratio $(2-2/k)d$ in dimension $d$.
Krumke et al.~\cite{krumke97} used the same ideas to prove that a variant of 
MM is a $(2-2/k)$-approximation algorithm; their argument also applies to MM.

\begin{theorem}
MC1x1 is a $(2-2/k)d$-approximation algorithm for minimizing the sum
of pairwise $L_1$ distances in a $d$-dimensional mesh.
\end{theorem}

\begin{proof}
Recall that MC1x1 minimizes the sum of
the selected points' shell numbers.
Let point $v$ be the center of the shells for the selected allocation
and let $\sigma$ be the sum of the shell numbers for points of $\MC$.
First, we bound $|\MC|$ in terms of $\sigma$.
The total distance from $v$ to each point of $\MC$ is at most
$\sigma d$ since a point in shell $i$ is at most $id$ steps from
$v$.
Thus, $|\MC| \leq (k-1)\sigma d$ since this is the distance if all paths
are routed through $v$. 

Now we bound $|\OPT|$ in terms of $\sigma$.
For this, we use the concept of a star, which is a set of points with one
identified as its center.
The length of a star is the total distance between the center
and its other points.
The smallest star with $k$ points has length at least $\sigma$ since a
point distance $i$ from the star's center is in the $i^\text{th}$ shell
around that center.
Thus, the total distance from one point of $\OPT$ to the others is at
least $\sigma$.
Since summing the lengths of stars of $\OPT$ with each point as the
center counts the distance between each pair of points twice,
$|\OPT| \geq k\sigma/2$ and the lemma follows by combining our bounds.
\qed
\end{proof}

\subsection{Fast Algorithm for $k=3$}

\begin{theorem}
Let $P$ be a set of $n$ points in the plane.
The subset of $P$ of size 3 with minimum total pairwise $L_1$ distance
can be found in $O(n \log n)$ time.
\end{theorem}

\begin{proof}
Let $S = \{s_0, s_1,s_2\}$ be a subset of $P$.
Label the $x$- and $y$-coordinates of a point $s \in S$ with
$(x_a,y_b)$ with $0 \leq a < 3$ and $0 \leq b < 3$ so that
$x_0 \leq x_1 \leq x_2$ and $y_0 \leq y_1 \leq y_2$.
The total pairwise $L_1$ distance of $S$ is $2(x_2-x_0) + 2(y_2-y_0)$.
Consider the smallest Steiner star of $S$, which has center $(x_1,y_1)$.
Its length is $(x_2-x_0) + (y_2-y_0)$.
Since the total pairwise distance and length of the smallest Steiner
star are constant multiples of each other, the 
subset of size 3 having minimum total pairwise distance also has the
smallest Steiner star.

Let $c$ be the center of the smallest Steiner star of 3 points of $P$.
By the discussion above, the three points having this Steiner star
also have minimum total pairwise distance.
These points are the three closest points to $c$ or
there would have been a smaller Steiner star.
Therefore, these points correspond to a cell on the order-3 Voronoi
diagram of $P$. 
Since this diagram can be found in $O(n \log n)$ time \cite{l-knnvd-82}, 
the theorem follows.
\qed
\end{proof}

%\vspace*{-4mm}
\section{PTAS for Two Dimensions}
\label{sec:ptas}

Let $w(S,T)$  be the sum of all the distances from points in
$S$ to points in $T$. Let  $w_x(S,T)$ and  $w_y(S,T)$ be the sum
of $x$- and $y$- distances from points in $S$ to
points in $T$, respectively. So  $w(S,T) = w_x(S,T) + w_y(S,T)$.
Let $w(S) = w(S,S)$, $w_x(S) = w_x(S,S)$, and $w_y(S) = w_y(S,S)$.
We call $w(S)$ the {\em weight} of $S$.

Let $S = \{s_0, s_1, \ldots, s_{k-1}\}$ be a minimum-weight subset
of $P$, where $k$ is an integer greater than 1. We label the
$x$- and $y$-coordinates of a point $s \in S$ by some $(x_a,y_b)$
with $0 \leq a < k$ and $0 \leq b < k$ such that $x_0 \leq x_1
\leq \ldots \leq x_{k-1}$ and $y_0 \leq y_1 \leq \ldots \leq
y_{k-1}$. (Note that in general, $a\neq b$ for a point
$s=(x_a,y_b)$.) We can derive the following equations:
\[ w_x(S) ~=~   (k-1)(x_{k-1}-x_0) + (k-3)(x_{k-2}-x_1) +~ \ldots ~   \]
and
\[ w_y(S) ~=~   (k-1)(y_{k-1}-y_0) + (k-3)(y_{k-2}-y_1) +~ \ldots ~   \]
We show that there is a polynomial-time approximation scheme
(PTAS), i.e., for any fixed positive $m=1/\eps$, there is a polynomial
approximation algorithm that finds a solution within
$(1+\eps)$ of the optimum.

\begin{figure}[htb]
   \begin{center}
   \input{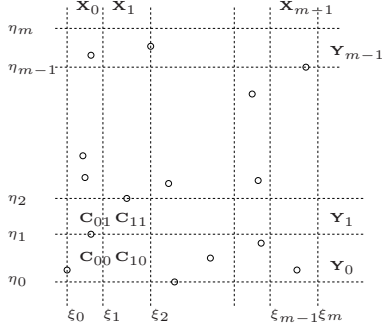}
   \caption{Dividing the point set into horizontal and vertical strips.}
   \label{fi:PTAS}
   \end{center}
\end{figure}

The basic idea is similar to the one used by Fekete and Meijer~\cite{fm-mdgmwc-04}
to select a set of points maximizing the overall
distance: We find (by enumeration) a subdivision of an
optimal solution into $m\times m$ rectangular cells $C_{ij}$, each
containing a specific number $k_{ij}$ of selected
points.  The points from each cell $C_{ij}$ are selected in a way
that minimizes the total distance to all other cells except
for the $m-1$ cells in the same ``horizontal'' strip or the $m-1$
cells in the same ``vertical'' strip. As it turns
out, this can be done in a way that the total neglected distance
within the strips is bounded by a small fraction of the weight of
an optimal solution, yielding the desired approximation property.
See Figure \ref{fi:PTAS} for the setup.

%\vspace*{-5mm}

%\vspace*{-8mm}
For ease of presentation, we assume that $k$ is a multiple of $m$
and $m > 2$. Approximation algorithms for other values of $k$ can
be constructed in a similar fashion.
Consider a division
of the plane by a set of $m+1$ $x$-coordinates
$\xi_{0}\leq\xi_{1}\leq\ldots\leq\xi_{m}$. Let $X_i:=\{p=(x,y)
\mid \xi_{i}\leq x\leq \xi_{i+1}\}$
be the
vertical strip between coordinates $\xi_i$ and $\xi_{i+1}$. By
enumeration of possible values of $\xi_{0},\ldots,\xi_{m}$ we may
assume that each of the $m$ strips $X_i$ contains precisely $k/m$ points
of an optimal solution.
(A small perturbation does not change
optimality or approximation properties of solutions.
Thus, without loss of generality, we assume that no pair of points share
either $x$-coordinate or $y$-coordinate.)

In a similar manner, assume we know $m+1$ $y$-coordinates
$\eta_0 \leq \eta_1 \leq \ldots \leq \eta_{m}$ so that
an optimal solution has precisely $k/m$ points in each
horizontal strip $Y_i:=\{p=(x,y) \mid \eta_i\leq y\leq \eta_{i+1}\}$.

Let $C_{ij}:=X_i\cap Y_j$, and let $k_{ij}$ be the number of
points in $\OPT$ that are chosen from $C_{ij}$.
Since for all $i,j \in \{1, 2, \ldots, m\}$,
$$\sum_{0\leq l < m} k_{lj}= \sum_{0 \leq l < m} k_{il}=k/m,$$
we may assume
by enumeration over the $O(k^m)$ possible partitions of $k/m$ into
$m$ pieces that we know all the numbers $k_{ij}$.

\begin{figure}[htb]
   \begin{center}
   \input{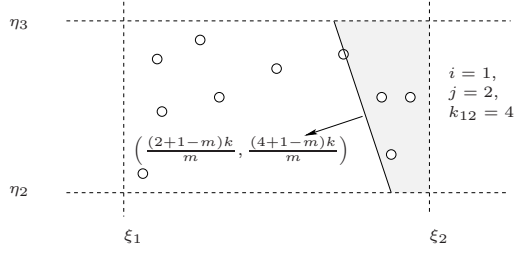}
   \caption{Selecting points in cell $C_{12}$.  }
   \label{fi:onecell}
   \end{center}
\end{figure}

Finally, define the vector $\nabla_{ij}:=
((2i+1-m)k/m,(2j+1-m)k/m)$. Our approximation algorithm is as
follows: from each cell $C_{ij}$, choose $k_{ij}$ points that
are minimum in direction $\nabla_{ij}$, i.e., select points $p =
(x,y)$ for which $(x(2i+1-m)k/m,y(2j+1-m)k/m)$ is minimum. For an
illustration, see Figure \ref{fi:onecell}.

It can be shown that selecting points of $C_{ij}$ this way
minimizes the sum of $x$-distances to points not in $X_i$
and the sum of $y$-distances to points not in $Y_j$.
Technical details are described in the following.
We summarize:

%\vspace*{-3mm}
\begin{theorem}
\label{th:ptas}
The problem of selecting a subset of minimum total $L_1$ distance
for a set of points in $\Re^2$ allows a PTAS.
\end{theorem}

%\vspace*{-10mm}

\subsection*{Correctness of the PTAS}
\label{sec:app.ptas}

Let \HEU\ be the point set selected by the algorithm described in
Section~\ref{sec:ptas}.
It is clear that $\HEU$ can be computed in polynomial time. We
will proceed by a series of lemmas to determine how well $w(\HEU)$
approximates $w(\OPT)$. In the following, we consider the
distances involving points from a particular cell $C_{ij}$. Let
$\HEU_{ij}$ be the set of $k_{ij}$ points that are selected from
$C_{ij}$ by the heuristic, and let $\OPT_{ij}$ be a set of
$k_{ij}$ points of an optimal solution that are attributed to
$C_{ij}$. Let $\HEUip$, $\OPTip$, $\HEUpj$ and $\OPTpj$ be the set
of $k/m$ points selected from $X_i$ and $Y_j$ by the heuristic and
an optimal algorithm respectively. Finally
$\ol{\HEU}_{i\bullet}:=\HEU\setminus\HEUip$, $\ol{\HEU}_{\bullet
j}:=\HEU\setminus\HEUpj$,
$\ol{\OPT}_{i\bullet}:=\OPT\setminus\OPTip$ and
$\ol{\OPT}_{\bullet j}:=\OPT\setminus\OPTpj$.

For the rest of the notation notice that

\begin{eqnarray*}
w(HEU) & = & \sum_{i,j} {[w_x(\HEUij,\notHEUip) + w_y(\HEUij,\notHEUpj)]}\\
         &+& \sum_{i}   {w_x(\HEUip)} + \sum_{j}   {w_y(\HEUpj)}.
\end{eqnarray*}

We first show that the first part is smaller that $w(\OPT)$. We
then show that the second and third part are small fractions of
$w(HEU)$.

\begin{lemma}
\label{le:external}
\begin{eqnarray*}
&&   w_x(\HEUij,\notHEUip) + w_y(\HEUij,\notHEUpj)\\
   &\leq&  w_x(\OPTij,\notOPTip) + w_y(\OPTij,\notOPTpj).
\end{eqnarray*}
\end{lemma}

\proof Consider a point $p \in \OPT_{ij} \setminus \HEU_{ij}$.
We will replace it with an arbitrary point
$p'\in \HEU_{ij}\setminus\OPT_{ij}$ that
was chosen by the heuristic instead of $p$.
Let $p-p' = h = (h_x,h_y)$.
When replacing $p'$ in $\HEU$ by $p$, we increase the
$x$-distance to the $ik/m$ points left of $C_{ij}$ by $h_x$, while
decreasing the $x$-distance to $(m-i-1)k/m$ points right of
$C_{ij}$ by $h_x$. In the balance, this yields a change of
$((2i+1-m)k/m)h_x$. Similarly, we get a change of
$((2j+1-m)k/m)h_y$ for the $y$-coordinates.  Since $p'$ was chosen
to minimize the inner product $\langle p' , \nabla_{ij}\rangle $ we
know that the inner product $\langle h,\nabla_{ij}\rangle \geq 0$,
so the overall change of distances is positive.

Performing these replacements for all points in $\HEU\setminus\OPT$,
we can transform $\HEU$ to $\OPT$,
while increasing the sum of distances
$w_x(\HEUij,\notHEUip) + w_y(\HEUij,\notHEUpj)$ to the sum
$w_x(\OPTij,\notOPTip) + w_y(\OPTij,\notOPTpj)$.\qed

\begin{corollary}
\label{co:external}
$$\sum_{i,j} {w_x(\HEUij,\notHEUip) + w_y(\HEUij,\notHEUpj)} \leq w(\OPT).$$
\end{corollary}

\bigskip
In the following two lemmas we show that $$\sum_i w_x(\HEUip)$$ is a
small fraction of $w(\HEU)$. Analogous proofs can be given for
$$\sum_j w_y(\HEUpj).$$

\begin{lemma}
\label{le:middlestrips}
\begin{eqnarray*}
   \sum_{0<i<m-1} w_x(\HEUip) ~\leq~ \frac{w_x(\HEU)}{m-2}.
\end{eqnarray*}
\end{lemma}

\proof Let $\delta_i = \xi_{i+1} - \xi_i$. Since $i(m-i-1)\geq
m-2$ for $0<i<m-1$, we have 
\[ w_x(\HEUip) \leq \frac{k^2}{m^2} \delta_i
               \leq \frac{ik}{m} \frac{(m-i-1)k}{m} \delta_i ~\frac{1}{m-2} \]
for $0 < i < m-1$.
Since $\HEU$ has $ik/m$ and $(m-i-1)k/m$ points to the left of
$\xi_i$ and right of $\xi_{i+1}$ respectively, we have
%\[
$$w_x(\HEU) \geq \sum_{0<i<m-1} \frac{ik}{m} \frac{(m-i-1)k}{m} \delta_i$$% \]
so
%\[
$$\sum_{0<i<m-1} w_x(\HEUip)  \leq \frac{1}{m-2} w_x(\HEU).$$% \]
\qed

\begin{lemma}
\label{le:endstrips} For $i = 0$ and $i=m-1$ we have
\[ w_x(\HEUip) ~\leq~ \frac{w_x(\HEU)}{m-1}. \]
\end{lemma}

\proof Without loss of generality assume $i=0$. Let $x_0, x_1,
\cdots , x_{(k/m)-1}$ be the $x$-coordinates of the points
$p_0,p_1,\ldots,p_{(k/m)-1}$ in $\HEUop$. So
\begin{eqnarray*}
w_x(\HEUop) &=&\left(\frac{k}{m}-1\right)\left(x_{\frac{k}{m}-1}-x_0\right)
             + \left(\frac{k}{m}-3\right)\left(x_{\frac{k}{m}-2}-x_1\right)
             +~\ldots\\
         &\leq&\left(\frac{k}{m}-1\right)(\xi_1-x_0)
             + \left(\frac{k}{m}-3\right)(\xi_1-x_1) +~\ldots\\
%         &\leq&\frac{k}{m} (\xi_1-x_0) + \frac{k}{m}(\xi_1-x_1) +~\ldots\\
         &\leq&\frac{k}{m} (\xi_1-x_0) + \frac{k}{m}(\xi_1-x_1) +~\ldots
            ~+ \frac{k}{m} \left(\xi_1-x_{\frac{k}{m}-1}\right).\\
\end{eqnarray*}
Since $\xi_1 - x_j \leq x - x_j$ where $0 \leq j < k/m$ and $x$ is
the $x$-coordinate of any point in $\notHEUop$ and since there are
$(m-1)k/m$ points in $\notHEUop$, we have
\[ \xi_1 - x_j < \frac{m}{(m-1)k} w_x (p_j,\notHEUop) \]
so
\begin{eqnarray*}
w_x(\HEUop) &\leq& \frac{k}{m} \frac{m}{(m-1)k}
                   \sum_{0 \leq i < \frac{k}{m}} w_x (p_i,\notHEUop)\\
            &\leq& \frac{1}{m-1}
                   \sum_{0 \leq i < \frac{k}{m}} w_x (p_i,\notHEUop)\\
               &=& \frac{1}{m-1} w_x (\HEUop, \notHEUop)\\
            &\leq& \frac{1}{m-1} w_x (\HEU).\\
\end{eqnarray*}
\qed

Combining the three lemmas, we get the claimed result and the proof
of Theorem~2.

\begin{eqnarray*}
w(\HEU) &=&\sum_{i,j} {w_x(\HEUij,\notHEUip) + w_y(\HEUij,\notHEUpj)}\\
        && + \sum_{i} {w_x(\HEUip}) + \sum_{j} {w_y(\HEUpj})\\
    &\leq&w(OPT) + \frac{1}{m-2} (w_x(\HEU) +  w_y(\HEU))\\
        && + \frac{2}{m-1} (w_x(\HEU) + w_y(\HEU))\\
       &=&w(OPT) + \frac{1}{m-2} w(\HEU) + \frac{2}{m-1} w(\HEU).\\
\end{eqnarray*}
So
\[ w(\HEU) \left(1 -  \frac{1}{m-2} - \frac{2}{m-1}\right)  ~\leq~ w(OPT).        \]

%\vspace*{-12mm}
\section{Higher-Dimensional Spaces}
\label{sec:high.d}

Using the same techniques,
we also generalize our results to higher
dimensions. We start by describing the performance of \ALG.

%\vspace*{-2mm}
\subsection{\boldmath $\left(2-\frac{1}{2d}\right)$-Approximation}
%\vspace*{-2mm}
As in two-dimensional space, \ALG\ enumerates over
the $O(n^d)$ possible medians.
For each median, it constructs a candidate solution of the  $k$ closest points.

%\vspace*{-3mm}
\begin{lemma}
\label{le:lower.d}
\ALG\ is not better than a $2-1/(2d)$ approximation.
\end{lemma}

%\vspace*{-5mm}
\begin{proof}
We construct an example based on the cross-polytope in $d$ dimensions, i.e.,
the $d$-dimensional $L_1$ unit ball.
Let $\eps>0$.
Denote the origin with $O$ and the $i^\text{th}$ unit vector with $e_i$.
The example has $k/2$ points at $O$ and $O+e_1$.
In addition, there are $k/(4d)$ points at $O-(1-\eps)e_1$,
$O+(2-\eps)e_1$, $O\pm(1-\eps)e_i$ for $i=2,\ldots,d$, and
$O+e_1\pm(1-\eps)e_i$ for $i=2,\ldots,d$.
\ALG\ does best with $O$ or $O+e_1$ as median, giving
a total distance of $(k^2/4)\left(2-1/(2d)\right)(1+\Theta(\eps))$.
Optimal is the points at $O$ and $O+e_1$,
giving a total distance
of $k^2/4$.
\qed
\end{proof}

Establishing a matching upper bound can be done analogously to Section~\ref{sec:two-approximation}.
Lemma~\ref{le:attain} holds
for general dimensions.
The rest is based on the following lemma, which is a
higher-dimensional version of Claim 5 in the proof of Theorem~\ref{th:7/4}:

\begin{lemma}
\label{le:freedom}
Worst-case arrangements for \ALG\ can be assumed to have all points
at positions $(0,\ldots,0)$ and $\pm e_i$, where $e_i$ is the
$i$th unit vector.
\end{lemma}

\sketch{
Consider a worst-case arrangement
within the cross-polytope centered at the origin with radius 1.
Local moves consist of continuous changes in point coordinates,
performed in a way that preserves the number of coordinate values.
This means that to move a point having a coordinate value different
from $0,1,-1$, then all other points sharing
that coordinate value are moved
to keep the identical coordinates the same, analogous to
the proof of Theorem~\ref{th:7/4}.
%Lemma~\ref{le:fore}. 

Note that under these moves, the functions OPT and MM are locally
linear, so the ratio of MM and OPT is locally constant, strictly
increasing,
or strictly
decreasing.
If a move decreases the ratio, the opposite move increases it,
contradicting the assumption that the arrangement is worst-case.

If the ratio is locally constant during a move, it will continue to
be extremal until an event occurs,
i.e., when the number of coordinate identities between points increases,
or the number of point coordinates at $0,1,-1$ increase.
While there are points with coordinates different from $0,1,-1$,
there is always a move that decreases the total degrees of freedom,
until all $dn$ degrees of freedom have been eliminated.
Thus, we can always reach an arrangement with
point coordinates values from the set $\{0,1,-1\}$. These leaves
the origin and the $2d$ positions $\pm e_i$
as only positions within the cross-polytope.
}

The restricted set of arrangements can
be evaluated with symmetry to yield

\begin{theorem}
\label{th:7/4.d}
For points lying in $d$-dimensional space, \MM\ is a $2-1/2d$-approxi\-mation algorithm,
which is tight.
\end{theorem}

\subsection{PTAS for General Dimensions}

\begin{theorem}
\label{th:ptas.d}
For any fixed $d$, the problem of selecting a subset of minimum total $L_1$ distance
for a set of points in $\Re^d$ allows a PTAS.
\end{theorem}

%\vspace*{-3mm}
\sketch{
For $m=\Theta(1/\eps)$, we subdivide the set of $n$ points
with $d(m+1)$ axis-aligned hyperplanes, such that $(m+1)$ are normal for each
coordinate direction. Moreover, any set of $(m+1)$ hyperplanes normal
to the same coordinate axis is assumed to subdivide the optimal solution
into $k/m$ equal subsets, called {\em slices}. Enumeration of all
possible structures of this type yields a total of
$n^{m}$ choices of hyperplanes in each coordinate, for a total
of $n^{md}$ possible choices.
For each choice, we have a total of $m^d$ cells,
each containing between $0$ and $k$ points; thus, there
are $O(m^{kd})$ different distributions of cardinalities to
the different cells.
As in the two-dimensional case, each cell
picks the assigned number of points
extremal in its gradient direction.

It is easily seen that for each coordinate $x_i$, the above choice
minimizes the total sum of $x_i$-distances between points not in the
same $x_i$-slice. The remaining technical part (showing that
the sum of distances within slices are small compared to the distances
between different slices) is analogous to the details described for the
two-dimensional case and omitted.
}

\section{Experiments}
\label{sec:exp}

The work discussed so far is motivated by the
allocation of a single job. In the following, we examine how well
our algorithms allocate streams of jobs; now the set of free
processors available for each job depends on previous allocations.

To understand the interaction between the quality of an individual
allocation and the quality of future allocations, we ran a simulation
involving pairs of algorithms.  One algorithm, the {\em
situation algorithm}, places each job.  This determines the free
processors available for the next job.  Each allocation decision
serves as an input to the other algorithm, the {\em decision
algorithm}.  Each entry in Table~\ref{pairwise-exp-table} represents
the average sum of pairwise distances for the decision algorithm with
processor availability determined by the situation algorithm.

Our simulation used the algorithms MC1x1, MM, MM+Inc, and HilbertBF.
MM+Inc uses local improvement on the allocation of MM,
replacing an allocated processor with an excluded
processor that improves average pairwise distance until it reaches a
local minimum.  HilbertBF is the 1-dimensional strategy of
Leung et al.~\cite{leung02a} used on Cplant.  The
simulation used the LLNL Cray T3D trace from the
Parallel Workloads Archive~\cite{workloads}.  This trace has
21323 jobs run on a machine with 256 processors, treated as a $16
\times 16$ mesh in the simulation.

\begin{table}
{\small
\begin{center}
\begin{tabular}{|l|r|r|r|r|} \hline
Situation & \multicolumn{4}{c|}{Decision Algorithm} \\ \cline{2-5}
Algorithm & MC1x1 & MM & MM+Inc & HilbertBF \\ \hline
MC1x1 &     5256 & 5218 & 5207 & 5432 \\ \hline
MM &        5323 & 5285 & 5276 & 5531 \\ \hline
MM+Inc &    5319 & 5281 & 5269 & 5495 \\ \hline
HilbertBF & 5090 & 5059 & 5046 & 5207 \\ \hline
\end{tabular}
\end{center}
\caption{Average sum of pairwise distances when the decision algorithm
makes allocations with input provided by the situation algorithm.}
\label{pairwise-exp-table}
}
\end{table}

%\vspace*{-10mm}
In each row, the algorithms are ranked in the order MM+Inc, MM, MC1x1,
and HilbertBF.  This is consistent with the worst-case performance
bounds; MM is a 7/4-approxi\-mation, MC1x1 is a 4-approximation,
and HilbertBF has approximation ratio $\Omega(N)$ on an $N \times N$ mesh.

\section{Conclusions}
\label{sec:conc}

The algorithmic work described in this paper is one step toward
developing algorithms for scheduling mesh-connected network-limited
multiprocessors. We have given provably good algorithms to
allocate a single job. The next step is to study the allocation of
job sequences, a markedly different
algorithmic challenge.

The difference between making a single allocation and a sequence of
allocations is already illustrated by the diagonal entries in
Table~\ref{pairwise-exp-table}, where the free processors depend on
the same algorithm's previous decisions. These give the ranking (from best
to worst) HilbertBF, MC1x1, MM+Inc, and MM.  The locally better
decisions of MM+Inc seem to paint the algorithm into a corner over
time. Figures~\ref{mc-figure}, \ref{fi:disc-town}, and
\ref{fi:cont-town} help explain why.
When starting on an empty grid, MC produces connected rectangular
shapes.
Locally, these shapes are slightly worse than the round shapes
produced by MM, but rectangles have better packing properties
because they avoid small patches of isolated grid nodes.

\addtocounter{footnote}{-2}

We confirmed this behavior 
over an entire trace using Procsimity~\cite{procsimity,windisch95b},
which simulates
messages moving through the network.  We ran the NASA Ames iPSC/860
trace\footnotemark \  from the Parallel Workloads Archive~\cite{workloads},
scaling down the number of processors for each job by a factor of 4.
This made the trace run on a machine with 32 processors, allowing us to
find the greedy placement that minimizes average pairwise
distance at that step.
For average job flow time, MC1x1 was
best, followed by MM, and then greedy.
We did not run MM+Inc in this simulation.
HilbertBF was much worse than all three of the
algorithms mentioned in part due to difficulties using it on a nonsquare mesh.

Based on these results and the work of Leung et al.~\cite{leung02a}, one of the
first allocators developed and licensed for the partially completed Red Storm
supercomputer uses a machine specific space-filling curve and a 1D bin-packing
technique.
We expect to have Red Storm implementations of a 3D version of MC1x1 and the
greedy heuristic (called MM) analyzed in this paper.

Thus, the online problem in an iterated scenario is the most interesting
open problem. 
We believe that
a natural attack may be to consider online packing of rectangular shapes
of given area. We plan to pursue this in future work.

\section*{Acknowledgments}
We thank Jens Mache for informative discussions on processor allocation.
We thank Moe Jette and Bill Nitzberg for providing the LLNL and NASA
Ames iPSC/860 traces, respectively, to the Parallel Workloads Archive.
Michael Bender was partially supported by 
Sandia and NSF Grants EIA-0112849 and CCR-0208670.
David Bunde was partially supported by Sandia and NSF grant CCR 0093348.
S\'andor Fekete was partially supported by DFG grants
FE 407/7 and FE 407/8.
Henk Meijer was partially supported by NSERC.
Sandia is a multipurpose laboratory operated by Sandia Corporation, a
Lockheed-Martin Company, for the United States Department of Energy under
contract DE-AC04-94AL85000.

\addtocounter{footnote}{1}

{\footnotesize
\bibliography{pairwise,techreport,refs,new-refs}
}

\end{document}